\documentclass[twocolumn,showpacs,pre,amsmath,amssymb,floatfix]{revtex4}
\usepackage{dcolumn}
\usepackage{amsmath,amssymb}
\usepackage{bm}
\usepackage{epsfig}
\usepackage{graphics}
\newcommand{\bd}{\bm}

\begin{document}

\title{Effective average action based approach to correlation functions at finite momenta}
\author{N. Hasselmann}
\affiliation{
Max-Planck-Institute for Solid State Research, Heisenbergstr.~1, D-70569 Stuttgart, Germany
}
\date{\today}
\begin{abstract}
  We present a truncation scheme of the effective average action approach of the
  nonperturbative renormalization group which allows for an accurate
  description of the critical regime as well as of correlation functions at
  finite momenta. The truncation is a natural modification of the standard
  derivative expansion which includes both all local correlations and
  two-point and four-point irreducible correlations to all orders in the
  derivatives. We discuss schemes for both the symmetric and the symmetry
  broken phase of the $O(N)$ model and present results for $D=3$. All
  approximations are done directly in the effective average action rather than in the
  flow equations of irreducible vertices.
  The approach
  is numerically relatively easy to implement and yields good results for
  all $N$ both for the critical exponents as well as for the momentum
  dependence of the two-point function.

\end{abstract}
\maketitle

The nonperturbative renormalization group (NPRG) technique is based on an exact flow equation
of the effective average action (or generating functional of irreducible
vertices) \cite{Wetterich93,Morris94} 
and has been applied to a large variety of systems, see e.g. 
Refs.~\cite{Berges02,Delamotte07,Metzner11,Kopietz10} for
reviews. It proved
especially useful when applied to critical phenomena where often even relatively
simple truncation schemes yield an accurate description of the critical
region \cite{Berges02,Delamotte07}. 
 While the exact flow
equation of the effective average action can almost never be solved, 
it allows for novel nonperturbative
approximation techniques. One successful approximation
strategy is the derivative expansion, where the effective average action is expanded consistently
to a given order in spatial derivatives, but no truncation is made in the
power of the fields. The derivative expansion has been applied with success
to $O(N)$ models \cite{Berges02,Canet03,Litim11}. 

The derivative expansion  allows strict control
over the symmetries of the studied models since all approximations are done in
the effective action which is expanded in invariants of the model. This approach
automatically yields flow equations which are both closed and further obey
the symmetry of the original model.
In contrast, if  a direct
field expansion of the effective action is employed and approximations are done at the
level of the flow equations of irreducible vertices, 
the invariance of the action is not guaranteed. In general, approximations at the level of
flow equations of vertices are therefore more difficult to control. On the other
hand, the derivative expansion can only access the asymptotic small momentum regime
of the theory and it cannot be applied to calculate correlation functions at finite
momenta.

Recently, several approximation schemes which allow to calculate correlation functions
at finite momenta were developed \cite{Ledowski04,Blaizot06,Sinner08,Benitez09}. 
The most sophistacted of these is the scheme presented in 
Refs.~\cite{Blaizot06,Benitez09} (BMW scheme) which is based on an approximate solution of
the exact flow equation of the two-point vertex in presence of a background field.
All of these approaches rely however on approximations at the
level of flow equations for irreducible vertices. 
 Instead, here we want 
to develop a scheme in which all approximations are done directly at the level
of the effective average action.
This allows for a  transparent
calculation
of the momentum dependence 
where the full symmetry of the model is always obeyed by the flow
equations.
While controlled, such a scheme does however not neccessarily guarantee more 
accurate results for the critical exponents.
The scheme we discuss below
is a  natural modification
of the usual derivative expansion and is based on a 
local potential which is supplemented by a  momentum dependent
potential 
which accounts for nonlocal correlations
up to the four-point vertex. Similar schemes were previously used
to calculate the one-particle spectral function of Bose condensates 
\cite{Sinner09,Sinner10} (see also \cite{Dupuis09,Eichler09})  and
also the thermal fluctuations of crystalline membranes such as graphene 
\cite{Braghin10} (see also \cite{Kownacki09}).
Here, we extend the scheme to include, besides the nonlocal terms, 
the full local potential and test
it on the $O(N)$ model, see e.~g. Ref.~\cite{Pelissetto02} for a recent
summary of
results on the $O(N)$ model.
 We develop two schemes, a nonlocal
potential approximation (NLPA) for the ordered state and a 
NLPA for the symmetric state. Both
schemes allow for an investigation of the critical region.
We first introduce in Sec.~\ref{sec:NLPA} the NLPA approach for both
the ordered state and the symmetric state. We present results
from a numerical solution of the flow equations in Sec.~\ref{sec:results},
with results for the critical exponents presented in subsection
\ref{sec:resultsa}. In subsection \ref{sec:resultsb} we assess the quality
of the approach in the finite momentum regime. We conclude in Sec.~\ref{sec:conclusions}.

\section{Nonlocal potential approximation} 
\label{sec:NLPA}
We begin with an approximation scheme for the ordered state, where
we explicitly incorporate a finite order parameter into the invariant
effective action.

\subsection{NLPA for the ordered state}

Starting point of the NLPA approach
is an effective average action which consists of both a 
nonlocal potential term,
which is restricted to second order in the invariant densities and is characterized by
the coupling function $u_\Lambda(k)$, and a local potential term 
$U_\Lambda(\rho-\rho_\Lambda^0)$
which may be an arbitrary function of $\rho-\rho_\Lambda^0$, where
$\rho_\Lambda^0$ is the 
(cutoff dependent)
order parameter density and $\rho=\bd{\varphi}^2/2$, where ${\bd \varphi}$ is
a field with $N$-components. To avoid double counting of
correlations, we define $u_\Lambda(k)$ to be completely nonlocal with $u_\Lambda(0)=0$.
We furthermore keep the full momentum dependence $\sigma_\Lambda(k)$ of the
quadratic term in the action,
and thus approximate the effective average action, after subtraction of the
noninteracting
contribution $(1/2)\int_k G_{0,\Lambda}^{-1}(k)  \bd{\varphi}_{\bd k} \cdot
{\bd \varphi}_{- {\bd k}}$, as
\begin{align}
\Gamma_\Lambda[\varphi]&= \frac{1}{2} \int_k \big[  \sigma_\Lambda(k)
\bd{\varphi}_{\bd k} \cdot
{\bd \varphi}_{-{\bd k}} +
 u_\Lambda(k)\Delta\rho_{\bd k} \Delta\rho_{-{\bd k}} \big] \nonumber \\
& \qquad + \int_x
U_\Lambda(\rho-\rho_\Lambda^0) \, ,
\label{eq:action}
\end{align}
where $\Delta\rho_{\bd k}$ is the Fourier transform of $\rho({\bd x})-\rho_\Lambda^0$.
We use the notation $\int_k =\int d^D k/(2 \pi)^D$ and $\int_x = \int d^D x$
for integrals over momenta and integrals over coordinate space, respectively.
$U_\Lambda(\tau)$ can, for finite cutoff $\Lambda$, be expanded in $\tau$,
\begin{equation}
U_\Lambda(\tau)=\sum_n \frac{1}{n!} U_\Lambda^{(n)} \tau^n \, ,
\end{equation}
with $\tau=\rho-\rho_\Lambda^0$ or, for the symmetric scheme discussed in
subsection \ref{sec:nlpasym}, $\tau=\rho$.
Note that the effective action (\ref{eq:action}) 
does not contain all terms of a complete derivative approximation to order $q^2$, in which
the derivative term of the action would also include an expansion to all powers of 
$\rho-\rho_\Lambda^0$. In the present scheme, one could easily improve
upon the action  (\ref{eq:action})  by including e.g. also additional terms  
which can
be parametrized by only one momentum. One obvious extension would
be to include a term
$\int_{x,y} [\rho({\bd x})-\rho_{\Lambda}^0]^2 [\rho({\bd y})-\rho_\Lambda^0]
\kappa({\bd x}-{\bd y})$, 
with some function $\kappa({\bd x})$ 
which would be a  generalization of 
the  $(\partial_\mu\rho)^2 (\rho-\rho_\Lambda^0)$  
term encountered in a derivative expansion. Such an extension is both
straightforward and numerically feasible.

At the same time, the present approach goes well beyond the derivative 
expansion in that it includes
the full momentum dependence in the first two terms of Eq.~(\ref{eq:action}). 
As in the derivative expansion, the effective average action obeys 
the full $O(N)$ invariance throughout the entire flow.
To determine the flow of $U_\Lambda$, we can use the standard
technique \cite{Wetterich93}
and evaluate the flow of $\Gamma_\Lambda[\bar{{\bd \varphi}}]$
for a homogeneous ($x$-independent) field $\bar{\bd \varphi}$ such that
$V^{-1}\Gamma_\Lambda[ \bar{\bd \varphi}]=U_\Lambda(\bar{\rho} -\rho_\Lambda^0)$
with $\bar{\rho}=\bar{\bd \varphi}^2/2$ and where $V$ is the volume.
We shall now assume that $N\geq 2$, so that there is at least one
gapless transverse mode. The flow of the local potential is then 
given by \cite{Wetterich93}
\begin{align}
  \partial_\Lambda U_\Lambda(\bar{\rho}-\rho_\Lambda^0)
  &=\frac{1}{2}\int_k \partial_\Lambda R_\Lambda(k)\big\{
  \bar{G}_{\Lambda,\parallel}(k,\bar{\rho}) \nonumber \\ &\qquad +
  (N-1)\bar{G}_{\Lambda,\perp}(k,\bar{\rho}) \big\} \, ,
  \label{eq:flowU}
\end{align}
where 
\begin{subequations}
  \begin{align}
    \bar{G}_{\Lambda,\perp}^{-1}(k,\bar{\rho})&=
    \sigma_\Lambda(k)+
    U^\prime_\Lambda(\bar{\rho}-\rho_\Lambda^0)+G_{0,\Lambda}^{-1}(k) \, ,
    \label{eq:Gperprho}
     \\
    \bar{G}_{\Lambda,\parallel}^{-1}(k,\bar{\rho})&= \sigma_\Lambda(k)+
    2\bar{\rho} [u_\Lambda(k)+U^{\prime\prime}_\Lambda(\bar{\rho}-\rho_\Lambda^0)]
    \nonumber \\ & \qquad
    +
    U^\prime_\Lambda(\bar{\rho}-\rho_\Lambda^0)+G_{0,\Lambda}^{-1}(k) \, .
    \label{eq:Gparallelrho}
  \end{align}
\end{subequations}
Here, the cutoff regulated noninteracting Green's function is
\begin{equation}
  G_{0,\Lambda}^{-1}(k)=k^2 + R_\Lambda(k) \, ,
\end{equation}
and $R_\Lambda(k)$ is a regulator for small momenta with $k\lesssim \Lambda$.
The only difference of Eq.~(\ref{eq:flowU})
to the standard form used in a derivative expansion
of $\Gamma_\Lambda$ is the presence of the full functions $u_\Lambda(k)$
and $\sigma_\Lambda(k)$ 
in Eqs.~(\ref{eq:Gperprho}) and (\ref{eq:Gparallelrho})
rather than just their leading terms of a 
$k$-expansion.
To determine the flows of $\sigma_\Lambda(k)$ and
$u_\Lambda(k)$ we invoke a field expansion of 
$\Gamma_\Lambda[\varphi]$ in terms of 
$\Delta\varphi_k^a=\varphi_k^a-\varphi_\Lambda^0 \delta_{a1} \delta_{k,0}$ 
with $\rho_\Lambda^0=(\varphi_\Lambda^0)^2/2$.
Here we have, without
loss of generality, assumed an order parameter field $\varphi_\Lambda^0$
which is directed in the $a=1$ direction of the internal space.
To determine the flows of $\sigma_\Lambda$ and $u_\Lambda$, we need
the lowest order irreducible vertices (up to four-point), which have the
form 

\begin{subequations}
  \begin{align}
    \Gamma_{\Lambda,ab}^{(2)}({\bd k},-{\bd k})&=\delta_{ab}\sigma_\Lambda (k) + 
    2 \delta_{a1}\delta_{b1}  \rho_\Lambda^0 
    \tilde{u}_\Lambda(k) \, ,
    \\
    \Gamma_{\Lambda,abc}^{(3)}({\bd k}_1,{\bd k}_2,{\bd k}_3) &= \varphi_\Lambda^0 
    \Big[\delta_{a1} \delta_{bc} \tilde{u}_\Lambda(k_1)
    +\delta_{b1}\delta_{ac} \tilde{u}_\Lambda(k_2)\nonumber \\
    & \quad
    +\delta_{c1}\delta_{ab} \tilde{u}_\Lambda(k_3) \Big] 
    + (\varphi_\Lambda^0)^{3}U_\Lambda^{(3)} 
    \delta_{a1}\delta_{b1}\delta_{c1} \, ,
    \\
    \Gamma_{\Lambda,abcd}^{(4)}({\bd k}_1\dots {\bd k}_4)&=\delta_{ab}\delta_{cd}
    \tilde{u}_\Lambda(k_{12})+\delta_{ac}\delta_{bd}
    \tilde{u}_\Lambda(k_{13}) \nonumber \\
    & \quad + 
    \delta_{ad}\delta_{bc} \tilde{u}_\Lambda(k_{14})
    + 2\rho_\Lambda^0 U_\Lambda^{(3)} 
    \Big[\delta_{ab}\delta_{cd} \nonumber \\
    & \quad
    \times (\delta_{a1}+\delta_{c1}) + 
    \delta_{ac}\delta_{bd}(\delta_{a1}+\delta_{b1})
    \nonumber \\
    & \quad +  \delta_{ad}\delta_{bc}(\delta_{a1}+\delta_{b1}) \Big] 
    \nonumber \\
    & \quad 
    +4(\rho_\Lambda^0)^2 U_\Lambda^{(4)} 
    \delta_{a1}\delta_{b1}\delta_{c1}\delta_{d1} \, ,
  \end{align}
\end{subequations}
where we defined $\tilde{u}_\Lambda(k)=u_\Lambda(k)+U_\Lambda^{(2)}$ and 
$k_{ij}=|{\bd k}_i+{\bd k}_j|$.
The flow of the order parameter  follows from the requirement
that $\partial_\Lambda\Gamma_\Lambda^{(1)}=0$. This yields \cite{Schuetz06} 
\begin{align}
  \partial_\Lambda \rho_\Lambda^0 &=\frac{-1}{2 \tilde{u}_\Lambda(0)}
  \int_q\Big\{ \big[
  \tilde{u}_\Lambda(0)+2\tilde{u}_\Lambda(q)+2\rho_\Lambda^0 U_\Lambda^{(3)} \big]
  \dot{G}_{\Lambda,\parallel}(q) \nonumber
  \\ & \qquad\qquad\qquad
  +(N-1) \tilde{u}_\Lambda(0) \dot{G}_{\Lambda,\perp}(q) \Big\} \, ,
  \label{eq:flowrho}
\end{align}
where $G_{\Lambda,\alpha}(k)=\bar{G}_{\Lambda,\alpha}(k,\rho_\Lambda^0)$ for 
$\alpha=\perp,\parallel$ and
$\dot{G}_{\Lambda,\alpha}(k)=- G_{\Lambda,\alpha}^2(k) \partial_\Lambda R_\Lambda(k)$.
The flow of $\sigma_\Lambda(k)$ follows from the flow of $\Gamma_{\Lambda,\perp}^{(2)}(k)
=\Gamma_{\Lambda,aa}^{(2)}(k,-k)$
where $a\neq 1$ is a direction transverse to the order parameter field.
\begin{align}
  \partial_\Lambda \sigma_\Lambda(k) &=
  \int_q \Big\{  \dot{G}_\perp(q) \tilde{u}_\Lambda(q^\prime)-
  \dot{G}_\parallel(q)\tilde{u}_\Lambda(q)\Big \} \nonumber \\
  & \quad -2 \rho_\Lambda^0 \int_q \Big\{\dot{G}_\parallel(q^\prime)
  G_\perp(q) \tilde{u}_\Lambda^2(q^\prime) \nonumber \\
  & \qquad \qquad +
  \dot{G}_\perp(q^\prime) G_\parallel(q)\tilde{u}_\Lambda^2(q) \Big\} \, ,
  \label{eq:flowsigma}
\end{align}
and we defined $q^\prime=|\bd{k}+\bd{q}|$.
The flow equation of $\tilde{u}_\Lambda(k)$ can be obtained from the flow of 
$\Gamma_{\Lambda,\parallel}^{(2)}(k)=\Gamma_{\Lambda, 11}^{(2)}(k,-k)$ 
which reads
\begin{align}
  \partial_\Lambda \Gamma_{\Lambda,\parallel}^{(2)}(k)&=
  \frac{1}{2}\int_q \Big\{(N-1)\dot{G}_{\Lambda,\perp}(q)\big[\tilde{u}_\Lambda(0)
  +2 \rho_\Lambda^0 U_\Lambda^{(3)}\big]
  \nonumber \\
  & \qquad \qquad
  + \dot{G}_{\Lambda,\parallel}(q)\big[\tilde{u}_\Lambda(0)+2\tilde{u}_\Lambda(q^\prime) 
\nonumber \\
  & \qquad \qquad \qquad
  +12 \rho_\Lambda^0 U_\Lambda^{(3)}
  +4 (\rho_\Lambda^0)^2 U_\Lambda^{(4)}\big] \Big\} \nonumber \\
  & 
  -2\rho_\Lambda^0\int_q\Big\{ (N-1)\dot{G}_{\Lambda,\perp}(q^\prime)G_{\Lambda,\perp}(q)\tilde{u}_\Lambda^2(k)
    \nonumber \\
  &  \qquad \qquad
  +\dot{G}_{\Lambda,\parallel}(q^\prime)G_{\Lambda,\parallel}(q) \big[\tilde{u}_\Lambda(q) +\tilde{u}_\Lambda(q^\prime)
  \nonumber \\
  &  \qquad \qquad \qquad
 +\tilde{u}_\Lambda(k)+2 \rho_\Lambda^0 U_\Lambda^{(3)} \big]^2
  \Big\}   
  \nonumber \\
  &
  +\big[ \tilde{u}_\Lambda(0)+2  \tilde{u}_\Lambda(k) +2 \rho_\Lambda^0 U_\Lambda^{(3)}\big] \partial_\Lambda \rho_\Lambda^0 \, .
  \label{eq:flowparallel}
\end{align}

\begin{table*}[ht]
  \caption{Values for the anomalous dimension $\eta$ for various
    $N$ and $D=3$
    from different
    approaches. The columns correspond to the symmetric NLPA, the  NLPA  
for the ordered phase, results from the background field scheme (BMW) 
    \cite{Benitez09}, the first order derivative expansion (DE),  field theory (FT),
    variational perturbation theory (VPT) \cite{Kleinert99} and Monte Carlo (MC).
    \label{Tab:eta}}
  \begin{ruledtabular}
    \begin{tabular}{l l l l l l l l }
      N & sym. NLPA & ord. NLPA  & BMW & DE & FT & VPT & MC \\
      \\
      0 & 0.042 & & 0.034 & 0.039 \cite{Gersdorff01}& 0.0272(3)
      \cite{Pogorelov08} & 0.031(1) & 0.0303(3) \cite{Grassberger97} 
      \\
      1& 0.042 &  & 0.039 & 0.0443 \cite{Canet03} 
      & 0.0318(3) \cite{Pogorelov08} &0.034(7)& 0.03627(1) \cite{Hasenbusch10} 
      \\
      2& 0.041(5) & 0.049 & 0.041 & 0.049 \cite{Gersdorff01} 
      & 0.0334(2) \cite{Pogorelov08} &0.035(6)& 0.0381(2) \cite{Campostrini06}
      \\
      3& 0.040 & 0.046 & 0.040 & 0.049 \cite{Gersdorff01}
      & 0.0333(3) \cite{Pogorelov08}&0.035(0)& 0.0375(5) \cite{Campostrini02} \\
      4& 0.038 & 0.042 & 0.038 & 0.047 \cite{Gersdorff01}& 0.0350(45)
      \cite{Guida98} &0.031& 0.0365(10) \cite{Hasenbusch01}  
      \\    
      10& 0.026 & 0.024(5) & 0.022 & 0.028 \cite{Gersdorff01} &0.024 \cite{Antonenko95}& 0.0216 &  \\
    \end{tabular}
  \end{ruledtabular}
\end{table*}

Combining Eqs.~(\ref{eq:flowrho}), (\ref{eq:flowsigma}), 
(\ref{eq:flowparallel}), and keeping in mind that 
$\Gamma_{\Lambda,\parallel}^{(2)}(k)=\sigma_\Lambda(k)+2\rho_\Lambda^0
\tilde{u}_\Lambda(k)$, one finds
\begin{align}
  \partial_\Lambda \tilde{u}_\Lambda(k)& =\frac{1}{2\rho_\Lambda^0}
  \int_q \big[\dot{G}_{\Lambda,\parallel}(q)-\dot{G}_{\Lambda,\perp}(q) \big]
  \tilde{u}_\Lambda(q^\prime)
  \nonumber \\ &
  + \int_q \dot{G}_{\Lambda,\parallel}(q) \Big\{2 U_\Lambda^{(3)}+\rho_\Lambda^0 U_\Lambda^{(4)}
  \nonumber \\ &  \qquad
  -U_\Lambda^{(3)} \big[\tilde{u}_\Lambda(q)+ \rho_\Lambda^0 
  U_\Lambda^{(3)}\big]/\tilde{u}_\Lambda(0) \Big\}
  \nonumber \\ & 
  -\int_q\Big\{ (N-1)\dot{G}_{\Lambda,\perp}(q^\prime)G_{\Lambda,\perp}(q)\tilde{u}_\Lambda^2(k)
    \nonumber \\
  &  \qquad \qquad
  +\dot{G}_{\Lambda,\parallel}(q^\prime)G_{\Lambda,\parallel}(q) \big[\tilde{u}_\Lambda(q) +\tilde{u}_\Lambda(q^\prime)
  \nonumber \\
  &  \qquad \qquad\qquad \qquad
  +\tilde{u}_\Lambda(k)+2 \rho_\Lambda^0 U_\Lambda^{(3)} \big]^2
  \Big\} 
  \nonumber \\
  &   
  +\int_q \Big\{\dot{G}_\parallel(q^\prime)
  G_\perp(q) \tilde{u}_\Lambda^2(q^\prime) \nonumber \\
  & \qquad \qquad +
  \dot{G}_\perp(q^\prime) G_\parallel(q)\tilde{u}_\Lambda^2(q) \Big\} \, .
  \label{eq:flowu}
\end{align}

This completes the derivation of the flow equations, which are uniquely determined by the
effective action (\ref{eq:action}). The flow Eqs.~(\ref{eq:flowU},\ref{eq:flowrho},\ref{eq:flowsigma},\ref{eq:flowu})
form a closed set which can be used to calculate the full momentum dependence of the self-energies
in a  controlled manner and the only approximation is the form of the effective
action as stated in Eq.~(\ref{eq:action}).
By construction, the approach reproduces exactly the correct structure
of the leading order perturbation theory, which is dominant
at large momenta. Also by construction,
it reproduces
the leading terms in a derivative expansion of both $u_\Lambda(k)$ and
$\sigma_\Lambda(k)$ to lowest order in the fields, 
which dominate the behavior in the infrared. The same is true also for
the symmetric scheme which we discuss below.

\begin{figure}[ht]
\includegraphics[angle=-90,width=7cm]{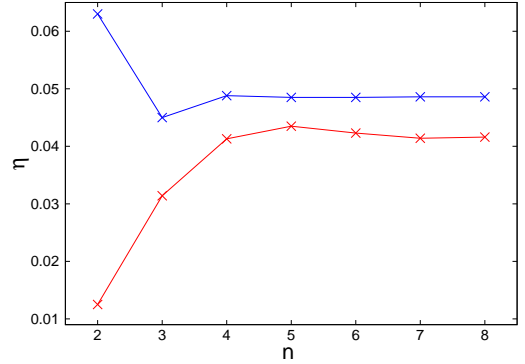}
\caption{Dependence of the anomalous dimension $\eta$ on
the order $n$ of the polynomial approximation of the local potential $U_\Lambda(y)=
\sum_{j=0}^n U_\Lambda^{(j)} y^j /j!$ for the symmetry broken phase
(upper curve) and the symmetric
scheme (lower curve) in which the fixed
point is approached from within the symmetric phase. Values shown are for
$N=2$ and $D=3$ .
} \label{fig:etavsp}
\end{figure}

\subsection{
 NLPA for the symmetric state}
\label{sec:nlpasym}
We now derive flow equations which are valid for the symmetric
phase, which are even simpler. In the NLPA 
for the symmetric state
the distance to the critical point is controlled by a 
mass term $r_\Lambda$ in the propagator which vanishes at criticality
in the limit $\Lambda\to 0$. We write 
the Ansatz for $\Gamma_\Lambda$ in the NLPA  as 
\begin{align}
  \Gamma_\Lambda[\varphi]&= \frac{1}{2} \int_k 
  \Big\{  [\sigma_\Lambda(k)+r_\Lambda] \bd{\varphi}_{\bd k} \cdot
  {\bd \varphi}_{-{\bd k}} +
  u_\Lambda(k) \rho_{\bd k} \rho_{-{\bd k}} \Big\} \nonumber \\
  & \qquad + \int_x
  U_\Lambda(\rho) \, ,
  \label{eq:actionsym}
\end{align}
where $\rho_k$ is the Fourier transform of $\rho({\bd x})={\bd \varphi}^2({\bd
  x})/2$ and where we put $U_\Lambda^{(1)}=0$ to avoid double counting of the
mass
term which is already accounted for by $r_\Lambda$.
The action (\ref{eq:actionsym}) yields again unique
flow equations for $r_\Lambda$ and the functions $U_\Lambda(\rho)$, 
$\sigma_\Lambda(k)$ and $u_\Lambda(k)$ 
which can be easily derived. 
We define the vertices now as expansion coefficients of 
$\Gamma_\Lambda$ around ${\bd \varphi}=0$. 
The flow for the two-point vertex 
$\Gamma_{\Lambda,ab}^{(2)}(k,-k)=\delta_{ab} \Sigma_\Lambda(k)$
 is then 
\begin{equation}
  \partial_\Lambda \Sigma_\Lambda(k)=\frac{1}{2} \int_q \dot{G}_\Lambda(q)
  \big[ 2 \tilde{u}_\Lambda(q^\prime)+N \tilde{u}_\Lambda(0) \big] \, ,
\end{equation}
where $\Sigma_\Lambda(k) = r_\Lambda + \sigma_\Lambda(k)$ with
$\sigma_\Lambda(0)=0$ and where 
$\dot{G}_\Lambda(q)=-G_\Lambda^2(q)\partial_\Lambda R_\Lambda(q)$
with $G_\Lambda^{-1}=G_{0,\Lambda}^{-1}+\Sigma_\Lambda(k)$.
 The flow of the two-point vertex is in the symmetric phase
not sufficient to extract also the flow of $\tilde{u}_\Lambda (k)=u(k)+U^{(2)}$ and
we must extract its flow from the four-point vertex. This yields
\begin{align}
  \partial_\Lambda \tilde{u}_\Lambda(k)&=\frac{4+N}{2} \int_q
  \dot{G}_\Lambda(q) U_\Lambda^{(3)} -\int_q \dot{G}_\Lambda(q) G_\Lambda(q^\prime)
  \Big\{ 
  \nonumber \\ & \quad
(N-1) \tilde{u}_\Lambda(k)^2+\big[\tilde{u}_\Lambda(k)+
  \tilde{u}_\Lambda(q^\prime)+\tilde{u}_\Lambda(q)\big]^2 \Big\}
\nonumber \\ & \quad
-\int_q \dot{G}_\Lambda(q) G_\Lambda(q) \Big\{ 
\nonumber \\ & \qquad
  [\tilde{u}_\Lambda(q^\prime)-\tilde{u}_\Lambda(q)]
[\tilde{u}_\Lambda(0)+2\tilde{u}_\Lambda(q)]\Big\} \, ,
\end{align}
with $q^\prime=|{\bd k}+{\bd q}|$.
The flow of the local potential $U_\Lambda(\rho)$ is given by 
Eq.~(\ref{eq:flowU}) 
with
$\rho_\Lambda^0=0$ and with $\sigma_\Lambda(k)$ 
replaced by $\sigma_\Lambda(k)+r_\Lambda$ 
in Eqs.~(\ref{eq:Gperprho}) and (\ref{eq:Gparallelrho}). 

\section{Results}
\label{sec:results}

We have solved the flow equations both in the symmetric phase and the symmetry
broken phase numerically for $D=3$ and different values of $N$. For $D=3$ the
field expansion of the local potential actually converges 
relatively fast \cite{Canet03} so that one can work with a finite 
order approximation of the local
potential. 
We have used an expansion of $U_\Lambda(\rho)$ up to eighth
order in $\rho$ in both the symmetric and the symmetry broken scheme
and have checked that the values of the anomalous dimension $\eta$ are already converged
at this level of truncation. The convergence can clearly be seen in 
Fig.~\ref{fig:etavsp} where we show the values of $\eta$ for
different maximal powers of $\rho$. All results presented below were
calculated with all terms up to order $\rho^8$.
 
\begin{table*}[ht]
  \caption{
Values for the anomalous dimension $\nu$ for various
    $N$ and $D=3$
    from different
    approaches. The columns correspond to the symmetric NLPA, the NLPA  
    for the ordered state, results from the background field scheme (BMW) 
    \cite{Benitez09}, the first order derivative expansion (DE),  field theory
    (FR), 
    variational perturbation theory (VPT) \cite{Kleinert99} and Monte Carlo (MC).
    \label{Tab:nu}}
  \begin{ruledtabular}
    \begin{tabular}{l l l l l l l l}
      N & sym. NLPA &ord. NLPA  & BMW & DE & FT &VPT& MC \\
      \\
      0& 0.58 & &0.589 & 0.590 \cite{Gersdorff01} & 0.5886(3) \cite{Pogorelov08}   & 0.5883 & 0.5872(5) \cite{Pelissetto07} \\
      1& 0.62 & &0.632 & 0.6307 \cite{Canet03}   & 0.6306(5)
      \cite{Pogorelov08}& 0.6305 & 0.63002(10) \cite{Hasenbusch10} \\
      2& 0.66 &0.68 &0.674 & 0.666 \cite{Gersdorff01}&
      0.6700(6) \cite{Pogorelov08} & 0.6710 & 0.6717(1) \cite{Campostrini06} \\
      3& 0.70 &0.72 &0.715 & 0.704 \cite{Gersdorff01} & 0.7060(7) \cite{Pogorelov08}& 0.7075 & 0.7112(5) \cite{Campostrini02}\\    
      4 & 0.74 &0.76 &0.754 & 0.739 \cite{Gersdorff01}& 0.741(6)
      \cite{Guida98}& 0.737 & 0.749(2) \cite{Hasenbusch01} \\
      10& 0.89 & 0.89& 0.889 & 0.859 \cite{Gersdorff01}& 0.859 \cite{Antonenko95} & 0.866 \\ 
    \end{tabular}
  \end{ruledtabular}
\end{table*}

For numerical stability we choose an exponential cutoff,
\begin{equation}
  R_\Lambda(q^2)=Z^{-1}_\Lambda \alpha \frac{q^2}{\exp(q^2/\Lambda^2)-1} \, ,
\end{equation}
where $Z_\Lambda^{-1}=1+\partial_{k^2} \sigma_\Lambda(k)|_{k=0}$ is the wavefunction
renormalization.
Usually the prefactor $\alpha$
is tuned in such a way as to extremize the critical exponents,
e.g. the anomalous dimension 
\begin{equation} \label{eq:eta}
\eta=\Lambda\partial_\Lambda \ln Z_\Lambda \, . 
\end{equation} This ensures a minimal sensitivity
of the results to small variations in $\alpha$ \cite{Canet03}. 
In the present scheme we do not observe an extremal value of $\eta$
as a function of $\alpha$. Instead, we observe a steady decrease
of $\eta$ when $\alpha$ is increased and a minimum which is only
reached asymptotically for large $\alpha$. For the symmetric scheme,
the dependence of $\eta$ on $\alpha$ is already essentially flat
for $\alpha\geq 5$ and we choose $\alpha=5$ for our analysis below.
Similarly, in the symmetry broken phase
only a small decrease of $\eta$ is detected on increasing
$\alpha$ from 1 to 2 and $\eta$ is then essentially unchanged up to $\alpha =
3$. We fixed $\alpha=2$ for the analysis below.

\subsection{Critical exponents $\eta$ and $\nu$}
\label{sec:resultsa}
The value of $\eta$ can be easily determined from the flow of the 
quantity $\sigma_\Lambda(k)$ and its low momentum structure via Eq.~(\ref{eq:eta}).
To determine the thermal exponent $\nu$, we use in the symmetric 
phase the value of 
the fully renormalized mass term  $r_*=\lim_{\Lambda \to 0} r_\Lambda$ which
scales as $r_*\simeq (r_{\Lambda_0} - r_c)^{2 \nu}$ where $r_c$ is the critical
value of the mass term at the initial cutoff scale $\Lambda_0$. 
Similarly, in the symmetry broken phase we analyse
the scaling of the order parameter $\rho_*=\lim_{\Lambda \to 0} \rho_\Lambda$ 
which scales as $\rho_*\simeq (\rho_{\Lambda_0}-\rho_c)^{2 \beta}$ where $\beta$
is the critical exponent of the order parameter and $\rho_c$ the critical
value of $\rho_\Lambda$  at the initial scale $\Lambda_0$.
From $\beta$ and $\eta$ we can extract
 $\nu$ via the hyperscaling relation $\nu=2\beta/(D-2+\eta)$.

In Tables \ref{Tab:eta} and \ref{Tab:nu} we show our results
for the critical exponents $\eta$ and $\nu$ and compare them with
results from various other approaches. 
Somewhat
surprisingly, and in contrast to what is observed in a standard derivative expansion, the
results for the critical exponents are generally better in the scheme where
 one approaches
the critical point from the symmetric side, where a field expansion
around $\rho=0$  rather than around a finite value $\rho^0$ is employed. 
As can be seen from Table \ref{Tab:eta}, the values for $\eta$ in the
symmetric scheme are, except for $N=0$, quite close to those
obtained within the BMW scheme of Ref.~\cite{Benitez09}.
For large $N$, it is known that $\eta$ behaves as $\eta=0.27/N$ \cite{Moshe03}, and
the result from the symmetric scheme for $N=10$ is already  close
to this value.

The results for the approach from the symmetry broken
phase are similar to those of the leading order derivative expansion
(where terms up to order ${\cal O}(q^2)$ are kept),
see Table \ref{Tab:eta}. A possible reason for the inferior accuracy of
 the scheme for the symmetry broken
phase compared with the accuracy of the symmetric scheme 
is that all nonlocal correlations are determined already from the 
two-point function whereas in the symmetric scheme the
nonlocal potential flow is determined from the four-point function.
Including further terms in the effective action is expected to 
improve also the results of the symmetry broken NLPA.

The results for the thermal exponent $\nu$ are similar in both schemes
and generally close to the most accurate MC results with deviations
never more than about 3\%. Our values are also close
to values from other approaches.

\subsection{Beyond the universal regime}
\label{sec:resultsb}
Both the schemes for the symmetric and 
the symmetry broken phase reproduce the logarithmic behavior
of the self-energy at large momenta 
$\Sigma(k)\simeq u_{\Lambda_0}^2 \ln (k/u_{\Lambda_0})$
which can
be derived from perturbation theory \cite{Baym01}. To assess the accuracy
of the calculated self-energy over the whole momentum regime a useful 
quantity is the small $u_{\Lambda_0}$ limit of
the one-dimensional integral ($\zeta(z)$ is the Riemann zeta function)
\begin{equation} 
c=\frac{128 }{3 \pi u_{\Lambda_0}} \zeta[3/2]^{-4/3} \int_0^\infty dq 
\frac{\Sigma(q)}{q^2+\Sigma(q)} \, ,
\label{Eq:defc}
\end{equation}
where $\Sigma(q)$ is the full self-energy at criticality, 
$\Gamma_{\Lambda,ab}^{(2)}(k,-k)=\delta_{ab} \Sigma_\Lambda (k)$ and
$\Sigma(k)=\lim_{\Lambda \to 0} \Sigma_\Lambda(k)$. 
The quantity $c$ is finite in the limit $u_{\Lambda_0}/\Lambda_0 \to 0$
and  has physical significance for $N=2$ where it relates to
the suppression of the critical temperature of the weakly interacting
Bose gas in $D=3$ dimensions \cite{Baym01}. The integral in Eq.~(\ref{Eq:defc}) is
dominated by contributions from the crossover regime $k\simeq u_{\Lambda_0}$
where the momentum dependence of the self-energy changes from the
perturbative $\ln (k)$ behavior at large momenta to the anomalous $k^{2-\eta}$
scaling at small momenta. The value of $c$ has been estimated
for different $N$ from Monte Carlo  simulations 
\cite{Arnold01,Kashurnikov01,Sun03} and has been used
to quantify the accuracy of various approaches 
\cite{Ledowski04,Kastening04,Benitez09}. To calculate $c$ we 
used a small initial value of $u_{\Lambda_0}$, 
$u_{\Lambda_0}/\Lambda_0=0.001$.
Again we find
generally better values for the symmetric scheme. For $N=2$, the value is
about $15 \%$ too high when compared to MC results and 
for $N=1$ the difference is slightly larger. For $N= 4$ the
difference is less than 5\%. In comparison with the BMW scheme the
differences are 8\% for $N=2$ and rapidly decrease for larger $N$,
see Table \ref{Table:c}. 

\begin{table*}[ht]
  \caption{Values for the quantity $c$ defined in Eq.~(\ref{Eq:defc}), from
both the symmetric NLPA and the NLPA for the ordered state
as well as from a perturbative FRG approach (PFRG), the background field scheme (BMW),
variational perturbation theory (VPT) and Monte Carlo (MC).}
\label{Table:c}
  \begin{ruledtabular}
    \begin{tabular}{l l l l l l l}
      N & sym. NLPA & ord. NLPA & PFRG \cite{Ledowski04} & BMW \cite{Benitez09} & VPT \cite{Kastening04} & MC \\
      \\  
      1& 1.38 &  & &1.15 &  1.07(10)& 1.09(9) \cite{Sun03} \\
      2& 1.49 & 1.60 & 1.23 &1.37 & 1.27(10)  & 1.29(5) \cite{Kashurnikov01} \\
       &          &        &  &     &               &  1.32(2) \cite{Arnold01} \\
      3& 1.59 & 1.72 & &1.50 & 1.43(11)  &  \\ 
      4& 1.68 & 1.82  && 1.63 & 1.54(11) & 1.60(10) \cite{Sun03} \\    
      10& 2.02 & 2.11 && 2.02 &  &   \\
    \end{tabular}
  \end{ruledtabular}
\end{table*}

\section{Conclusions}
\label{sec:conclusions}
We have presented a straightforward nonlocal potential approximation
which allows access to finite momentum properties of correlation
functions and also allows for an accurate calculation of critical
exponents. As in the derivative expansion, in the NLPA 
all truncations are done at the level of the effective
action, a property it 
shares with the derivative expansion. This allows for a strict control of the
symmetries of the underlying model and also allows for 
extensions of the approach. While the present approach includes both
terms of arbitrary powers in the fields (in the local term), the nonlocal
terms are restricted up to fourth order in the fields. 
In contrast, in
the BMW scheme  \cite{Benitez09} all vertices have a momentum dependence
which is however only approximately taken into account.
The present scheme can easily be
extended by including for example terms of the 
type $\rho({\bd x}) \rho({\bd y})^2 \kappa({\bd x}-{\bd y})$
which
would result in momentum dependent vertices with up to six legs
and
similar terms of higher order in the densities can of course also easily be 
constructed. In $D=3$ it might suffice to limit such terms only to a small maximal
power in $\rho$ to
get converged values 
for the critical exponents.
For each such additional term a new coupling function must be
introduced so the nonlocality in the scheme will always be restricted to a
finite order in the fields. 
 The computational cost of such an extension is
relatively modest, since one would still deal with the flow of a small
number
of one-parameter functions. In contrast, in the BMW
the flow must be analysed for a two point function which is defined on a two-dimensional
grid, one dimension each for the dependence on fields and momenta, which 
is numerically more difficult.

Even in the the simplest NLPA truncation analysed here, the results for both
the critical exponents and for the momentum dependence of the 
two-point function are already surprisingly good and the present scheme offers
a direct access to both universal and non-universal quantities. 
The present scheme is also certainly useful for more complex models
 where
even the local terms are restricted to a finite
order in the fields \cite{Braghin10,Sinner09,Sinner10,Dupuis09,Kownacki09,Eichler09}.

 We thank 
Pawel Jakubczyk,
Andreas Eberlein, and  Federico Benitez for discussions.
This work was supported by the DFG research group FOR 723.


\begin{thebibliography}{99}
%
\bibitem{Wetterich93} C. Wetterich, Phys. Lett. B {\bf 301}, 90 (1993). 
\bibitem{Morris94} T. R. Morris,
Int. J. Mod. Phys. A {\bf 9}, 2411 (1994).
%
\bibitem{Berges02} J.~Berges, N.~Tetradis, and C.~Wetterich, Phys.~Rep.~{\bf
363}, 223 (2002).
%
\bibitem{Delamotte07} B. Delamotte, e-print arXiv:cond-mat/0702365 (2007).
%
\bibitem{Metzner11} W.~Metzner, M.~Salmhofer, C.~Honerkamp, V.~Meden, and 
K.~Sch\"onhammer, Rev.~Mod.~Phys. {\bf 84}, 299 (2012).
%
\bibitem{Kopietz10} P. Kopietz, L. Bartosch, and F. Sch\"utz, {\em Introduction to
the Functional Renormalization Group}, (Springer, Berlin,2010).

\bibitem{Canet03} L. Canet, B. Delamotte, D. Mouhanna, and J. Vidal, Phys. Rev.
D {\bf 67}, 065004 (2003); Phys. Rev. B 68, 064421 (2003); L. Canet,
{\em ibid.} 71, 012418 (2005).

\bibitem{Litim11} D.~F.~Litim and D.~Zappal\'a, Phys.~Rev.~D {\bf 83}, 085009
  (2011).

\bibitem{Ledowski04} S. Ledowski, N. Hasselmann, and P. Kopietz, {Phys. Rev.} A {\bf 69}, 061601(R) (2004); N. Hasselmann, S.
 Ledowski, and P. Kopietz, 
{\em ibid. } {\bf 70}, 063621 (2004).

\bibitem{Blaizot06}
J.~P.~Blaizot, R.~M\'{e}ndez-Galain, and N.~Wschebor, {Phys.~Lett.~B} 
{\bf 632}, 571 (2006).
%
\bibitem{Benitez09}
F.~Benitez, J.~P.~Blaizot, H.~Chat\'{e}, B.~Delamotte, 
R.~M\'{e}ndez-Galain, and N.~Wschebor, {Phys.~Rev.~E} {\bf 85}, 026707 (2012);
{\em ibid.} {\bf 80}, 030103(R) (2009).


\bibitem{Sinner08} A. Sinner, N. Hasselmann, and P. Kopietz, 
{J. Phys.: Cond. Mat.} {\bf 20}, 075208 (2008).
%
\bibitem{Sinner09} A. Sinner, N. Hasselmann, and P. Kopietz, {Phys. Rev.
Lett.} {\bf 102}, 120601 (2009).
%
\bibitem{Sinner10} A. Sinner, N. Hasselmann, and P. Kopietz, {Phys. Rev.
B} {\bf 82}, 063632 (2010).
%
\bibitem{Dupuis09} N.~Dupuis, Phys. Rev. Lett. {\bf 102}, 190401 (2009); 
Phys. Rev. A {\bf 80}, 043627 (2009).
%
\bibitem{Eichler09} C. Eichler, N. Hasselmann, and P. Kopietz, Phys. Rev. E
{\bf 80}, 051129 (2009).
%

\bibitem{Braghin10} N.~Hasselmann and F.~Braghin, Phys.~Rev.~E {\bf 83},
  031137 (2011);
F.L. Braghin and N. Hasselmann, {Phys. Rev. B} {\bf 82},
  035407 (2010).


\bibitem{Kownacki09} J.-P.~Kownacki and D.~Mouhanna, {Phys. Rev. E} {\bf 79},
  040101(R) (2009); 
K. Essafi, J.-P.~Kownacki, and D.~Mouhanna, Phys. Rev. Lett. {\bf 106}, 128102
(2011).
%
\bibitem{Pelissetto02} A.~Pelissetto and E.~Vicari, Phys. Rep. {\bf 368}, 549 (2002).

\bibitem{Schuetz06} F. Sch\"utz and P. Kopietz, J. Phys. A {\bf 39}, 8205 (2006).
%

\bibitem{Kleinert99} H.~Kleinert, Phys.~Rev.~D {\bf 60}, 085001 (1999).

 \bibitem{Gersdorff01} G. v. Gersdorff and C. Wetterich, 
Phys. Rev. B {\bf     64},  054513 (2001).

\bibitem{Pogorelov08} A. A. Pogorelov and I. M. Suslov, J. Exp. Theor. Phys. 
{\bf 106}, 1118 (2008).

\bibitem{Grassberger97} P. Grassberger, P. Sutter, and L. Sch¨afer, J. Phys. A
  {\bf 30}, 7039 (1997).

\bibitem{Hasenbusch10} M. Hasenbusch, Phys. Rev. B {\bf 82}, 174433 (2010).

\bibitem{Campostrini06} M. Campostrini, M. Hasenbusch, A. Pelissetto, and E. Vicari,
Phys. Rev. B {\bf 74}, 144506 (2006).

\bibitem{Campostrini02} M. Campostrini, M. Hasenbusch, A. Pelissetto, P. Rossi, and
E. Vicari, Phys. Rev. B {\bf 65}, 144520 (2002).

\bibitem{Guida98} R. Guida and J. Zinn-Justin, J. Phys. A {\bf 31}, 8103
(1998).

\bibitem{Hasenbusch01} M. Hasenbusch, J. Phys. A 34, 8221 (2001).

\bibitem{Moshe03} M. Moshe and J. Zinn-Justin, Phys. Rep. {\bf 385}, 69
  (2003).

\bibitem{Antonenko95} S. A. Antonenko and A. I. Sokolov, Phys. Rev. E {\bf 51}, 1894
(1995).


\bibitem{Pelissetto07} A. Pelissetto and E. Vicari, J. Phys. A {\bf 40}, F539
  (2007).

\bibitem{Baym01} G. Baym, J.-P. Blaizot, M. Holzmann, F. Laloe, and
  D. Vautherin, Eur.~Phys.~J.~B {\bf 24}, 107 (2001).

\bibitem{Kastening04} B. M. Kastening, Phys. Rev. A {\bf 69}, 043613 (2004).

\bibitem{Sun03} X. Sun, Phys. Rev. E 67, 066702 (2003).


\bibitem{Kashurnikov01} V. A. Kashurnikov, N. V. Prokof\'ev, and
 B. V. Svistunov, Phys. Rev. Lett. {\bf 87}, 120402 (2001).

\bibitem{Arnold01}  P. Arnold and G. Moore, Phys. Rev. Lett. {\bf 87}, 120401
(2001).




\end{thebibliography}
\end{document}